\documentclass[reprint,aps,amsmath,amssymb,twoside,prd,showkeys,superscriptaddress]{revtex4-1}
\pdfoutput=1
\usepackage{verbatim}
\usepackage{comment}
\usepackage{graphicx}
\usepackage[T1]{fontenc}

\usepackage{epsfig}
\usepackage{bm}
\usepackage{amssymb}
\usepackage{float}
\usepackage{amsmath}
\usepackage{subfigure}
\usepackage{dcolumn}
\usepackage{cancel}
\usepackage[colorlinks]{hyperref}
\usepackage[usenames,dvipsnames]{color}
\hypersetup{
     breaklinks=true,
    pdfstartview={FitH},    % fits the width of the page to the window
    colorlinks=true,       % false: boxed links; true: colored links
    linkcolor=blue,          % color of internal links
    citecolor=red,        % color of links to bibliography
    filecolor=magenta,      % color of file links
    urlcolor=blue,           % color of external links
    anchorcolor=green,      % Color for anchor text
    linktocpage=true
}

\newcommand{\HCd}{\mathcal{H}}

\def\HCdt0{\tilde{\HCd}_{0}}

\newcommand\rf[1]{(\ref{eq:#1})}
\newcommand\lab[1]{\label{eq:#1}}
\newcommand\nonu{\nonumber}
\newcommand\br{\begin{eqnarray}}
\newcommand\er{\end{eqnarray}}
\newcommand\be{\begin{equation}}
\newcommand\ee{\end{equation}}

\newcommand\lb{\lbrack}
\newcommand\rb{\rbrack}

\newcommand\bc{\begin{center}}
\newcommand\ec{\end{center}}

\usepackage{verbatim}
\renewcommand\d{\delta}
\newcommand\eps{\epsilon}
\newcommand\vareps{\varepsilon}

\newcommand\h{\frac{1}{2}}
\renewcommand\k{\kappa}
\renewcommand\l{\lambda}
\renewcommand\L{\Lambda}
\newcommand\m{\mu}
\newcommand\n{\nu}
\newcommand\om{\omega}

\newcommand\vp{\varphi}
\renewcommand\P{\Phi}
\newcommand\pa{\partial}

\newcommand\pr{\prime}

\newcommand\z{\zeta}

\newcommand\wti{\widetilde}
\newcommand\cB{{\mathcal B}}

\newcommand\cF{{\mathcal F}}

\newcommand\cL{{\mathcal L}}
\newcommand\cM{{\mathcal M}}

\newcommand{\ct}[1]{\cite{#1}}

\newcommand\Udot{\stackrel{.}{U}}

\newcommand\phidot{\stackrel{.}{\phi}}
\newcommand\phiddot{\stackrel{..}{\phi}}

\newcommand\Psidot{\stackrel{.}{\Psi}}

\newcommand\adot{\stackrel{.}{a}}

\newcommand\Bdot{\stackrel{.}{B}}
\newcommand\Hdot{\stackrel{.}{H}}

\newcommand{\afffias}{Frankfurt Institute for Advanced Studies (FIAS), Ruth-Moufang-Strasse~1, 60438 Frankfurt am Main, Germany}

\newcommand{\affbgu}{Physics Department, Ben-Gurion University of the Negev, Beer-Sheva 84105, Israel}

\newcommand{\affbul}{Institute for Nuclear Research and Nuclear Energy, Bulgarian Academy
of Sciences, Sofia, Bulgaria}
\begin{document}
\title{$\Lambda$CDM as a  Noether Symmetry in Cosmology}
\author{David Benisty}
\email{benidav@post.bgu.ac.il}
\affiliation{\affbgu}\affiliation{\afffias}
\author{Eduardo I. Guendemlan}
\affiliation{\affbgu}\affiliation{\afffias}
\author{E. Nissimov}
\email{nissimov@inrne.bas.bg}
\affiliation{\affbul}
\author{S. Pacheva}
\email{svetlana@inrne.bas.bg}
\affiliation{\affbul}
\begin{abstract}
The standard $\Lambda$CDM model of cosmology is formulated as a simple
modified gravity coupled to a single scalar field (``darkon'') possessing a
non-trivial hidden nonlinear Noether symmetry. The main ingredient in the
construction is the use of the formalism of non-Riemannian spacetime 
volume-elements. The associated Noether
conserved current produces stress-energy tensor consisting of two additive parts 
-- dynamically generated dark energy and dark matter components non-interacting 
among themselves. Noether symmetry breaking via an additional scalar ``darkon'' 
potential introduces naturally an interaction between dark energy and dark matter.
The correspondence between the $\Lambda$CDM model and the present ``darkon'' 
Noether symmetry is exhibited up to linear order w.r.t. gravity-matter perturbations.
\end{abstract}
\maketitle
\section{Introduction}
The recent realization that the Universe expansion is accelerating 
\cite{Perlmutter:1998np,Riess:1998cb} has puzzled cosmologists to this day and has 
lead them to conjecture the existence of dark energy (in the form of a  non-zero
cosmological constant $\Lambda$) and cold dark matter 
(CDM) -- called $\Lambda$CDM cosmological model. 
Even though the $\Lambda$CDM model presents a good fit to 
the present observations, it has some conceptual problems 
\cite{Weinberg:2000yb,Peebles:2002gy} motivating us to explore other possibilities 
for the dark sector. One enticing possibility is a form of dynamical dark energy 
\cite{Ratra:1987rm,Wetterich:1994bg} in which the acceleration is induced by a scalar 
field, usually referred to as quintessence models 
\cite{Zlatev:1998tr,Chiba:1999ka,Barreiro:1999zs,Caldwell:1997ii,dePutter:2007ny,Tsujikawa:2013fta,Babichev:2018twg,Kehayias:2019gir,Oikonomou:2019muq,Chakraborty:2019swx,Chervon:2019sey,Bento:2002ps,Benisty:2020vvm,Benisty:2020xqm,Benisty:2020qta}. 
Dark matter can also be described via a scalar field as weakly-interacting 
massive particles (WIMPs) -- still undiscovered 
at colliders and dark matter detection experiments. Models for dark matter can also 
be based on other kinds of scalar fields. This is for example the case of fuzzy 
dark matter \cite{Hu:2000ke}. Interaction between dark matter and dark energy was 
considered in many cases  
\cite{Arevalo:2016epc,Anagnostopoulos:2017iao,Saridakis:2018unr,Anagnostopoulos:2018jdq,Vagnozzi:2019ezj,Vasak:2019nmy,Benisty:2016ybt}. Interacting scenarios prove to be efficient in alleviating the known tension of modern cosmology   \cite{Yang:2018euj,Yang:2018uae,Guo:2018ans,Benisty:2018gzx,Kumar:2019wfs,Agrawal:2019lmo,Benisty:2019vej,Benisty:2019jqz,DiValentino:2019jae,DiValentino:2019ffd,Pan:2019gop,Ikeda:2019ckp,Yang:2019qza,Yang:2019uzo,Yang:2020uga,Benisty:2020otr}.

In order to provide a unified description of 
dark energy and dark matter through a simple scalar field one can use different 
extensions of the canonical scalar field action 
\cite{Leon:2013qh,Chamseddine:2013kea,Golovnev:2013jxa,Chamseddine:2014vna,Chaichian:2014qba,Matsumoto:2015wja,Myrzakulov:2015nqa,Cognola:2016gjy,dusty,Guendelman:2016kwj,Dutta:2017fjw,Benisty:2017lmt,Benisty:2017eqh,Casalino:2018tcd,Benisty:2018oyy,Benisty:2018qed,Anagnostopoulos:2019myt}. 
Ref.~\cite{dusty} uses the formalism of non-Riemannian spacetime volume-forms 
(NRVF -- see Section \ref{sec:NRVF} below) 
in addition to the canonical Riemannian volume-element $\sqrt{-g}$ defined by the
square-root of the determinant of the Riemannian metric. 
This NRVF construction yields a simple model of a  modified gravity coupled to a single
scalar field with two main features: (i) It dynamically generates
non-zero cosmological constant as a free integration constant not present in
the original model; (ii) It produces a non-trivial hidden nonlinear
Noether symmetry of the modified scalar field action, whose associated conserved 
Noether current yields the CDM part of the pertinent energy density. Thereby
the scalar field is called ``darkon'' and the associated nonlinear Noether
symmetry - ``darkon'' symmetry.

In the present paper we investigate the cosmological solutions of the above
``darkon'' model. We show that up to linear order of the metric and
``darkon'' field perturbations the hidden nonlinear ``darkon''  Noether symmetry yields 
energy density consisting of two separate dark energy and dark matter
contributions. Breaking of the Noether symmetry is introduced by an
additional ``darkon'' field potential leading to an interaction between dark
energy and dark matter components. The implications of the breaking of ``darkon''
Noether symmetry for a possible explanation of the cosmic tensions are briefly 
discussed.

The plan of the paper is as follows. Section \ref{sec:NRVF} briefly introduces the
main features of the NRVF formalism.
In Section \ref{sec:model} the basics of the ``darkon'' model are presented,
specifically the emergence and the role of the hidden nonlinear ``darkon''
Noether symmetry, including the dynamical generation of the dark matter
component of the energy density as a dust fluid flowing along geodesics. 
Section \ref{sec:cosmo} describes the homogeneous cosmological solution of
the unperturbed ``darkon'' model whereas in Section \ref{sec:pert} 
the perturbations of the latter are derived. In Section \ref{sec:Data} a plausible form of a $\Lambda$CDM Noether symmetry-breaking ``darkon''
potential is proposed and the corresponding solutions compared with some
observational data.
Finally, Section \ref{sec:conclusion} summarizes the results and discusses 
possible solutions to the cosmic tensions using the above formalism.

%%%%%%%%%%%%%%%%%%%%%%%%%%%%%%%%%%%%%%%%%%%%%%%%%%%%%%%%%%%%%%%%%%%%%%%%%%%%%%%%%%%
\section{The Essence of the Non-Riemannian Volume-Form formalism}
\label{sec:NRVF}

Volume-forms define generally covariant integration measures on differentiable 
manifolds (not necessarily Riemannian ones, so {\em no} metric is needed)
\cite{spivak}. They are given by nonsingular maximal-rank differential 
forms $\om$ (for definiteness we will consider the case of $D=4$ spacetime
dimensions):
\be
\int_{\cM} \om \bigl(\ldots\bigr) = \int_{\cM} dx^4\, \Omega \bigl(\ldots\bigr)
\;
\lab{omega-1}
\ee
with:
\br
\om = \frac{1}{4!}\om_{\m\n\k\l} dx^{\m}\wedge dx^{\n}\wedge dx^{\k}\wedge dx^{\l}
\nonu \\ 
\om_{\m\n\k\l} = - \vareps_{\m\n\k\l} \Omega,\\
\Omega = \frac{1}{4!}\vareps^{\m\n\k\l} \om_{\m\n\k\l} \;\nonu .
\lab{omega-2}
\er
The conventions for the alternating symbols $\vareps^{\m\n\k\l}$ and
$\vareps_{\m\n\k\l}$ are: $\vareps^{0123}=1$ and $\vareps_{0123}=-1$.
The volume-element density (integration measure density)
$\Omega$ transforms as scalar density under general coordinate reparametrizations.

In standard generally-covariant theories the Riemannian spacetime volume-form is 
defined through the tetrad canonical one-forms $e^A = e^A_\m dx^\m$ ($A=0,1,2,3$):
\be
%e^0 \wedge e^1 \wedge e^2 \wedge e^3 =
\om =  \det\Vert e^A_\m \Vert\,
dx^{0}\wedge dx^1 \wedge dx^2 \wedge dx^{3} \; ,
\lab{omega-riemannian-1}
\ee
which yields:
\be
\Omega = \det\Vert e^A_\m \Vert = \sqrt{-\det\Vert g_{\m\n}\Vert} \; .
\lab{omega-riemannian-2}
\ee
Instead of $\sqrt{-g} d^4 x$ we can employ another alternative {\em non-Riemannian} 
volume-element as in \rf{omega-1} given by a non-singular {\em exact} $4$-form 
$\om = d B$ where:
\br
B = \frac{1}{3!} B_{\m\n\k}
dx^{\m}\wedge dx^{\n}\wedge dx^{\k} . % \quad \longrightarrow \quad
\lab{Phi-4}
\er
Therefore, the corresponding non-Riemannian volume-element density
\be
\Omega \equiv \cF(B) =
\frac{1}{3!}\vareps^{\m\n\k\l}\, \pa_{\m} B_{\n\k\l}.
\lab{omega-nonriemannian}
\ee
is defined in terms of
the dual field-strength scalar density of an auxiliary rank 3 tensor gauge field 
$B_{\m\n\k}$. 

The systematic application of non-Riemannian volume-elements to construct
modified gravity-matter models was originally proposed in 
Refs.\cite{Guendelman:1996qy,Gronwald:1997ei,Guendelman:1999tb,Guendelman:1999qt,Guendelman:2007ph}, 
with a subsequent concise geometric formulation in \cite{susyssb-1,grav-bags}.
Let us particularly note the following important property of Lagrangian action terms 
involving (one or more independent) non-Riemannian volume-elements as in 
\rf{omega-nonriemannian}
%%%%%%%%%%%%%%%%%%%%%%%%
alongside with the canonical Riemannian volume element
%%%%%%%%%%%%%%%%%%%%%%%%
:
\be
% S = \int d^4 x \,\sum_j \cF(B^{(j)})\, \cL^{(j)} ({\rm other ~fields}) + \ldots
% begin-modified
S = \int d^4 x \,\sum_j \cF(B^{(j)})\, \cL^{(j)} 
+ \int d^4 x \,\sqrt{-g} \cL  \; .
% end-modified
\lab{NRVF-0}
\ee
The equations of motion of \rf{NRVF-0} with respect to the auxiliary tensor
gauge fields $B^{(j)}_{\m\n\k}$ according to \rf{omega-nonriemannian} imply:
\be
\pa_\m \cL^{(j)}({\rm other ~fields}) = 0 \quad \to \quad 
\cL^{(j)}({\rm other ~fields}) = \cM_j,
\lab{L-M}
\ee
where $\cM_j$ are {\em free integration constants} not present in the
original action \rf{NRVF-0}. 

The appearance of the free integration constants in \rf{L-M} plays instrumental role 
in the application of the NRVF formalism as a basis for constructing % a series of 
modified gravity-matter models describing unified dark energy and dark matter 
scenario \cite{dusty,dusty-2} (see also Section \ref{sec:model} below),
quintessential cosmological models with gravity-assisted and inflaton-assisted
dynamical suppression (in the ``early'' universe) or dynamical generation (in the
post-inflationary universe) of electroweak spontaneous symmetry
breaking and charge confinement \cite{grf-essay,varna-17,bpu-10}, as well as a novel 
mechanism for the supersymmetric Brout-Englert-Higgs effect (dynamical
spontaneous supersymmetry breaking) in supergravity \cite{susyssb-1}.
%%%%%%
For a systematic numerical study of some of the cosmological models proposed above on 
the basis of NRVF formalism, see  \cite{Staicova:2016pfd,Staicova:2018yrc}.

% The feature \rf{L-M} gives an elegant solution to the cosmological constant problem
% \cite{} or to be the basis for the unimodular gravity \cite{}.

%%%%%%%%%%%%%%%%%%%%%%%%%%%%%%%%%%%%%%%%%%%%%%%%%%%%%%%%%%%%%%%%%%%%%%%%%%%%

\section{Hidden Nonlinear Noether Symmetry}
\label{sec:model}

%%%%%%%%%%%%%%%%%%%%%%%%%%%%%%%
\subsection{``Darkon'' Model}

Our starting point is a modified gravity-matter model where the scalar field action 
consists of two terms -- one coupled to the standard Riemannian 
volume-element \rf{omega-riemannian-2} and a second one coupled to a 
non-canonical non-Riemannian one \rf{omega-nonriemannian} 
(using units with $16\pi G_{\rm Newton}=1$):
\be
S = \int d^4 x\, \Bigl\lb\sqrt{-g} \bigl(R + X-V_1(\phi)\bigr) 
+ \cF(B)\bigl( X-V_2(\phi)\bigr)\Bigr\rb \; ,
\lab{NRVF-1a}
\ee
where $R$ is the Ricci scalar, and $X$ is the kinetic term of a scalar field: 
\be
X \equiv - \h g^{\m\n} \pa_\m \phi \pa_\n \phi \; .
\lab{X-def}
\ee
%%%%%

The model \rf{NRVF-1a}, first considered in Refs.\cite{dusty,dusty-2}, is a
simple special case of the broad class of modified gravity-matter models
based on the NRVF formalism as in Eq.\rf{NRVF-0}.
%%%%%

We can equivalently reformulate the action \rf{NRVF-1a} as:
\br
S = \int d^4 x\, \sqrt{-g} \bigl(R - U(\phi)\bigr)
\nonu \\
+ \int d^4 x\, \bigl(\sqrt{-g}+\cF(B)\bigr)\bigl(X - V(\phi)\bigr)
\lab{NRVF-1}
\er
using the notations:
\be
V \equiv V_2 \;\; ,\;\; U \equiv V_1 - V_2 \; .
\lab{notat}
\ee
% There are three independent variables: the auxiliary field $B_{\mu\nu\rho}$ 
% that builds the modified measure, the scalar field $\phi$ which plays the 
% dynamical matter field and the metric of the space time $g_{\mu\nu}$.

\vspace{.1in}

%%%%%%%%%%%%%%%%%%%%%%%%%%  BEGIN COMMENT 1  %%%%%%%%%%%%%%%%%%%%%%%%%%%%%
Let us point out that the Riemannian metric $g_{\m\n}$ entering Eqs.\rf{NRVF-1a} or \rf{NRVF-1} is the physical "Einstein-frame" metric to which 
other generic matter field do couple, excluding the specific non-generic scalar field $\phi$ 
entering \rf{NRVF-1} which plays a special role in the present construction to 
describe dark matter (see next subsection). Thus, other generic fields $u$
which could be added, will appear in the action \rf{NRVF-1} as:
\br
S = \int d^4 x\, \sqrt{-g} \Bigl\lb R - \h g^{\m\n}\pa_\m u \pa_\n u + \ldots\Bigr\rb
\nonu \\
+ \int d^4 x\, \Bigl\lb \sqrt{-g} \bigl( X - V(\phi) - U(\phi)\bigr)
+\cF(B) \bigl(X- V(\phi)\bigr)\Bigr\rb \; .
\nonu \\
\phantom{aaaa}
\lab{NRVF-1b}
\er
Thus, the canonical energy-momentum tensor for the generic field $u$ (from the first line in \rf{NRVF-1b}) will be
obviously conserved and, therefore, $u$ will follow the standard minimally coupled equations w.r.t. the original 
metric $g_{\m\n}$, \textsl{i.e.}, which can be considered as the equivalence
principle for $u$ in \rf{NRVF-1b} is satisfied and there are no 5-th force effects with respect to these additional fields  $u$. We will latter address the issue of the fifth force for the effective dust that is generated by the modified measure theory.

The above case is in contrast w.r.t. other modified gravity models based on the 
non-Riemannian volume-form formalism 
\ct{Guendelman:1996qy,Gronwald:1997ei,Guendelman:1999tb,Guendelman:1999qt,Guendelman:2007ph,susyssb-1,grav-bags},
where the scalar curvature $R$ in the initial modified gravity action 
couples to certain non-Riemannian volume element
$\int d^4 x\,\cF(B) R + \dots$ (cf. \rf{omega-nonriemannian}-\rf{NRVF-0}), and where 
the physical Einstein-frame metric $g^{(EF)}_{\m\n}$ is {\em different} from the initial metric $g_{\m\n}$ -- it is obtained upon conformal transformation:
\begin{equation}
g_{\m\n} \to g^{(EF)}_{\m\n} = \frac{\cF(B)}{\sqrt{-g}} g_{\m\n}   .
\end{equation}
Variation of the action \rf{NRVF-1} w.r.t. auxiliary gauge field $B_{\m\n\k}$ 
inside $\cF(B)$ \rf{omega-nonriemannian} yields (cf. the general Eq.\rf{L-M}):
\be
\pa_\m \Bigl(X - V(\phi)\Bigr) = 0 \quad \to \quad X - V(\phi) = - 2M, 
\lab{L-const}
\ee
where $M$ is free integration constant not present in the original action
\rf{NRVF-1}.

The variation of \rf{NRVF-1} w.r.t. scalar field $\phi$ can be written in
the following suggestive form:
\br
\nabla_\m J^\m = - \sqrt{2X} U^\pr (\phi) \; ,
\lab{J-nonconserv} \\
J_\m \equiv - \bigl(1+\chi\bigr)\sqrt{2X} \pa_\m \phi \;\; ,\;\;
\chi \equiv \frac{\cF (B)}{\sqrt{-g}} \; .
\lab{J-current} 
\er
The dynamics of $\phi$ is entirely determined by
the dynamical constraint \rf{L-const}, completely independent of the potential 
$U(\phi)$. On the other hand, the $\phi$-equation of motion written in the form 
\rf{J-nonconserv} is in fact an equation determining the
dynamics of $\chi$.  

The energy-momentum tensor $T_{\m\n}$ in the Einstein equations following from
\rf{NRVF-1} ($R_{\m\n} - \h g_{\m\n}R = \h T_{\m\n}$), upon taking into account 
\rf{L-const} and \rf{J-current}, reads:
\be
T_{\m\n} = g_{\m\n}\bigl(-2M-U(\phi)\bigr) + (1+\chi)\pa_\m \phi \pa_\n \phi \;.
\lab{EM-tensor}
\ee
Both \rf{EM-tensor} and \rf{J-current} can be represented in a relativistic
hydrodynamical form for an ideal fluid:
\be
T_{\m\n} = \rho_0 u_\m u_\n + g_{\m\n} {\wti p}\;\;,\quad J_\m = \rho_0 u_\m
\lab{EM-hydro}
\ee
where $u_\m$ is the fluid velocity unit vector:
\be
u_\m \equiv - \frac{\pa_\m \phi}{\sqrt{2X}} \quad 
({\rm note} \; u^\m u_\m = -1\;) \; ;
\lab{fluid-velocity}
\ee
the energy density ${\wti \rho}$ and pressure ${\wti p}$ are given as:
% and $\rho_0 \equiv {\wti \rho} + {\wti p}$ with:
\be
{\wti \rho} = \rho_0 + 2M + U(\phi) \quad ,\quad {\wti p} = -2M - U(\phi)
\lab{fluid-rho-p}
\ee
with:
\be
\rho_0 \equiv (1+\chi) 2X = {\wti \rho} + {\wti p} \; .
\lab{rho-0}
\ee

Energy-momentum conservation $\nabla^\n T_{\m\n}= 0$ implies:
\be
%\nabla^\n T_{\m\n}= 0 \;\; \to \;\; {\rm Eq.\rf{J-nonconserv}} \;\; ,\;\;
\nabla^\m \bigl(\rho_0 u_\m\bigr) = - \sqrt{2X}\,U^\pr(\phi) \;\; 
\bigl({\rm Eq.\rf{J-nonconserv}}\bigr) \;\; , \;\;
u_{\n} \nabla^{\n} u_\m =0 \; , 
\lab{EM-conserve}
\ee
the last Eq.\rf{EM-conserve} meaning that the matter fluid flows along geodesics.
%%%%%%%%%%%%%%%%%%%%%%%%%%  BEGIN COMMENT 2  %%%%%%%%%%%%%%%%%%%%%%%%%%%%%

Notice now that the issue of the 5th force exists for the effective dust generated by the theory,
 and there will be $5^{th}$ force if the potential 
that appears in the first eq. in eq \rf{EM-conserve}, (or \rf{J-nonconserv}) is not flat, since the dust is not conserved then, which obviously mean 
an interaction between the scalar field and the dust (5th force), even though the four velocity satisfies 
the geodesic equation, ie the equivalence principle is satisfied (second eq, in eq \rf{EM-conserve}).
The particles obey the equivalence  principle, but the particle number is not conserved, the $5^{th}$ force manifests itself by particle creation or destruction (depending on the sign of the potential ) of our effective dust. Notice that in this case the departure from $\Lambda$CDM (departure from constant potential) is correlated with the appearance of $5^{th}$ force (because of the non constant potential). Notice that the Noether symmetry that we will discuss in the next section holds only if the potential $U$ is constant, so the Noether symmetry guarantees the absence of a fifth force. Notice that there are not very strong bounds in cosmology for a fifth force, although in the range of laboratory experiments there are. 

%%%%%%%%%%%%%%%%%%%%%%%%%%
\subsection{Hidden Nonlinear Noether Symmetry}

In Ref.\cite{dusty} a crucial property of the model \rf{NRVF-1} has been
uncovered for the special case with the potential $U(\phi)=0$:
\br
S^{(0)}= \int d^4 x\,\Bigl\lb \sqrt{-g} R 
+ \bigl(\sqrt{-g}+\cF(B)\bigr)\bigl(X - V(\phi)\bigr)\Bigr\rb \; .
\lab{NRVF-2}
\er
The variation with respect to the scalar field yields a {\em conserved current}
(cf. Eqs.\rf{J-nonconserv}-\rf{J-current}):
\be
\nabla^\m J_\m = 0 \;\;\; ,\;\; 
J_\m = - \bigl(1+\chi\bigr)\sqrt{2X} \pa_\m \phi = \rho_0 u_\m \;.
\lab{J-conserv}
\ee
$J_\m$ \rf{J-conserv} is  a genuine Noether conserved current of the action \rf{NRVF-2} 
corresponding to the following hidden strongly nonlinear symmetry transformations:
\br
\d_\eps \phi = \eps \sqrt{X},\quad \d_\eps g_{\m\n} = 0 \; ,
\nonu \\
\d_\eps \cB^\m = - \eps \frac{1}{2\sqrt{X}} \phi^{,\mu} 
\bigl(\P(B) + \sqrt{-g}\bigr)  \; ,
\lab{hidden-sym}
\er
with $\cB^\m \equiv \frac{1}{3!} \vareps^{\m\n\k\l} B_{\n\k\l}$. 
Under \rf{hidden-sym} the action \rf{NRVF-2} transforms as total
derivative of:
\be
\d_\eps S^{(0)} = \int d^4 x \, \pa_\m \bigl( L(\vp,X) \d_\eps \cB^\m \bigr).
\ee
The existence of the hidden Noether symmetry \rf{hidden-sym} of the action 
\rf{NRVF-2} {\em does not} depend on the specific form of the potential $V(\phi)$ 
in the  scalar field Lagrangian. The only requirement is that the kinetic term 
$X$ must be positive.

The hidden Noether symmetry \rf{hidden-sym} is valid also
for the action \rf{NRVF-1} in the particular case $U(\phi) = {\rm const}$.

The energy-momentum tensor corresponding to $S^{(0)}$ \rf{NRVF-2}, \textsl{i.e.},
Eq.\rf{EM-hydro} with \rf{fluid-rho-p} for $U(\phi)=0$, simplifies to:
\be
T^{(0)}_{\m\n} = \rho_0 u_\m u_\n - 2M g_{\m\n} \equiv \bigl(\rho +p\bigr) +
g_{\m\n} p,
\lab{EM-hydro-0}
\ee
with $\rho_0$ as in \rf{rho-0}.
Now the fluid tension $p=-2M$ is constant and negative, whereas the (total) fluid
energy density $\rho = \rho_0 + 2M$, so that $\rho_0$ 
\rf{fluid-rho-p} and $2M$ are the rest-mass and internal fluid energy
densities, respectively (for general definitions, see \textsl{e.g.}
\cite{rezzola-zanotti}).

The energy-momentum tensor \rf{EM-hydro-0} is an exact sum of two additive parts 
with the following interpretation of $\rho$ and $p$ in\rf{EM-hydro-0} according 
to the standard $\L$CDM model \cite{Lambda-CDM-1,Lambda-CDM-2,Lambda-CDM-3}:
\be
p = -2M = p_{\rm DM} + p_{\rm DE}\;\; ,\;\;
\rho = \rho_0 + 2M = \rho_{\rm DM} + \rho_{\rm DE} \; .
\lab{DE-DM-split}
\ee
Namely, taking into account \rf{J-conserv} and last Eq.\rf{EM-conserve} we have:

\begin{itemize}
\item
Dark energy part $\rho_{\rm DE}= - p_{\rm DE} = 2M$,
which arises due to the dynamical constraint on the scalar field
Lagrangian \rf{L-const}. 
\item
Dark matter part $p_{\rm DM} = 0$ and $\rho_{\rm DM} = \rho_0 \equiv \\ (1+\chi)2X$,
\textsl{i.e.}, dark matter appears as a dust-like fluid flowing along
geodesics and with conserved particle number density.
\end{itemize}

The above interpretation justifies the alias ``darkon'' for the scalar field $\phi$.
Let us specifically  emphasize that both dark energy and dark matter components of the
energy density have been {\em dynamically} generated thanks to the
non-Riemannian volume-element construction -- both due to the appearance of the
free integration constant $M$ and of the hidden nonlinear Noether symmetry.

On the other hand, when we start with the initial action \rf{NRVF-1} with
the addition of a Noether symmetry breaking potential
$U(\phi) \neq 0$, Eqs.\rf{EM-hydro}-\rf{fluid-rho-p} tell us that $U(\phi)$ triggers
an interaction (energy transfer) between the dark energy and dark matter
components due to the ``darkon'' $\phi$-dynamics:
\br
\rho_{\rm DE}=-p_{\rm DE} = 2M + U(\phi) \;,
\nonu \\
\rho_{DM} = \rho_0 \equiv \bigl(1+\chi\bigr) 2X
% \Bigl(1+\frac{\cF(B)}{\sqrt{-g}}\Bigr) \bigl(-g^{\m\n}\pa_\m\phi \pa_\n\phi\bigr)
% \Bigl(1+\frac{\cF(B)}{\sqrt{-g}}\Bigr)\,2\,\Bigl(V(\phi) - 2M\bigr)
\quad ,\quad p_{\rm DM} = 0 \; .
\lab{rho-p-DE-DM}
\er
Dark matter fluid is again dust-like fluid flowing along geodesics (second
Eq.\rf{EM-conserve}), however now because of the breakdown (first
Eq.\rf{EM-conserve} -- non-conservation of $J_\m$ \rf{J-current}) of the hidden 
nonlinear Noether symmetry the dark matter 
particle number density is not any more conserved. 

%%%%%%%%%%%%%%%%%%%%%%%%%%%%%%%%%%%%%%%%%%%%%%%%%%%%%%%%%%%%%%%%%%%%%%%%%%%
\section{Homogeneous Unperturbed Evolution}
\label{sec:cosmo}

Let us now perform a reduction of the action \rf{NRVF-1} to the FLRW
(Friedmann-LeMaitre-Robertson-Walker) metric:
\be
ds^2 = - dt^2 + a(t)^2 \d_{ij} dx^i dx^j 
\lab{FLRW-metric}
\ee
%with all fields being only $t$-dependent (after the variations the lapse
%function $N(t)$ is set to the gauge $N(t)=1$ -- gauge-fixing the residual 
%$t$-reparametrization invariance of \rf{FLRW-metric})):
%\br
%S_{\rm FLRW} = \int dt \Bigl\lb - 6 \frac{a \adot^2}{N} 
%+\bigl( Na^3 + \Bdot\bigr)\Bigl(\frac{\phidot^2}{2 N^2} - V(\phi)\Bigr)\Bigr\rb 
%\nonu \\
%+ \int dt\,Na^3 U(\phi) \; . \phantom{aaaaaaaaaaaaaaa}
%\lab{NRVF-1-FLRW}
%\er
Variation of \rf{NRVF-1}  w.r.t. $B$ yields the FLRW-reduced form of the dynamical constraint \rf{L-const}:
\be
\frac{d}{dt} \Bigl(\h\phidot^2 - V(\phi)\Bigr)=0 \quad \to \quad
\h\phidot^2 - V(\phi) = - 2M \; .
\lab{L-const-FLRW}
\ee
Taking time-derivative of \rf{L-const-FLRW} implies:
\be
\phiddot = V^{\pr}(\phi) \; ,
\lab{phiddot}
\ee
note the opposite sign in the ``force'' term on the r.h.s. of \rf{phiddot}.
According to \rf{L-const-FLRW} the solution for $\phi(t)$ reads:
\be
\int^{\phi(t)}_{\phi(0)} \frac{d\phi}{\sqrt{2\bigl(V(\phi) - 2M\bigr)}} = t
\; .
\lab{phi-sol}
\ee

The equation of motion of \rf{NRVF-1} w.r.t. $\phi$
is equivalent to the FLRW-reduction of \rf{J-nonconserv}, which amounts to
an equation for the dark matter energy density $\rho_0$:
\br
\Bigl(\frac{d}{dt} + 3H\Bigr) \rho_0 + \frac{d}{dt} U(\phi) = 0 \; ,
\lab{J-nonconserv-FLRW} \\
\rho_0 (t) \equiv (1+\chi ) \phidot^2 
= \frac{c_0}{a^3 (t)} - \frac{1}{a^3(t)} \int dt^\pr a^3(t^\pr) \Udot (t^\pr)
\; .
\lab{rho-0-FLRW}
\er
Here $c_0$ is an integration constant, $\Udot \equiv U^\pr(\phi)\phidot$, and \\
% $\chi \equiv \frac{\Bdot}{a^3}$ 
$\chi \equiv \Bdot\!\!/a^3$ is the FLRW-reduced form of the ratio of
volume-element densities $\chi \equiv \frac{\cF(B)}{\sqrt{-g}}$ (last Eq.\rf{J-current}).

In the case of $U(\phi)=0$ when the nonlinear ``darkon'' Noether symmetry is
intact Eqs.\rf{J-nonconserv-FLRW}-\rf{rho-0-FLRW} reduce to:
% in complete analogy with the 4-dimensional case \rf{J-conserv}, 
% is equivalent to the FLRW-reduction of Noether symmetry current conservation
% \rf{J-conserv}:
\be
\Bigl(\frac{d}{dt} + 3H\Bigr) \rho_0 = 0 \quad \to \quad 
\rho_0 \equiv (1+\chi) \phidot^2 = \frac{c_0}{a^3} \; .
% \phidot \left(3 H (\chi+1)+\dot{\chi}\right) + 2 (\chi+1)V^\pr (\phi)  
% + U^\pr(\phi) = 0  \; ,
\lab{J-conserv-FLRW}
\ee
where the Hubble parameter $H = \frac{\adot}{a}$. The last
Eq.\rf{J-conserv-FLRW} explicitly exhibits the dust-like nature of the ``darkon''
dark matter energy density $\rho_0$.

The Friedmann equations % -- variations of \rf{NRVF-1} w.r.t. to $N$ and $a$ -- 
read accordingly:
\br
6 H^2 = {\wti \rho} \quad ,\quad {\wti\rho} = \rho_0  + 2M + U(\phi)
\lab{fried-1}\\
\Hdot = - \frac{1}{4} \bigl({\wti\rho} + {\wti p}\bigr) \equiv - \frac{1}{4} \rho_0
\quad ,\quad {\wti p} = - 2M - U(\phi) \; , 
\lab{fried-2}
\er
where ${\wti\rho}$ and ${\wti p}$ are as in \rf{EM-hydro}-\rf{fluid-rho-p} and
$\rho_0$ is given now by the homogeneous solution \rf{rho-0-FLRW}. 

In the case of $U(\phi)=0$ when the nonlinear ``darkon'' Noether symmetry is
intact, taking into account \rf{J-conserv-FLRW}, Eqs.\rf{fried-1}-\rf{fried-2} 
simplify to:
\br
6 H^2 = \rho,\quad \rho = \rho_0  + 2M \equiv \frac{c_0}{a^3} + 2M \; ,
\lab{fried-1-0}\\
\Hdot = - \frac{1}{4} \bigl(\rho + p\bigr) \equiv - \frac{c_0}{4a^3},\quad p = - 2M \;. 
\lab{fried-2-0}
\er

For comparison with the observational data it is convenient to rewrite 
Eqs.\rf{phi-sol}-\rf{rho-0-FLRW} and \rf{fried-1} in terms of function w.r.t. 
red-shift variable $z$:
\be
1+z = \frac{a_0}{a(t)} \quad ,\quad 
\frac{d}{dt} = - (1+z)H(z) \frac{d}{dz} \; ,
\lab{z-def}
\ee
as follows:

\begin{itemize}
\item
Eq.\rf{phi-sol} is equivalent to introducing the ``darkon'' field
redefinition:
\be
\phi \;\to \; {\wti\phi} = \wti{\phi} (\phi) \quad ,\quad
\frac{\pa\wti{\phi}}{\pa\phi} = \Bigl\lb 2\bigl( V(\phi) - 2M\bigr)\Bigr\rb^{-1/2} \; ,
\lab{phi-tilde}
\ee
so that:
\be
\Bigl(\frac{d\wti{\phi}}{dt}\Bigr)^2 = 1 \quad \to \quad
\frac{d\wti{\phi}}{dz} = - \frac{1}{(1+z)H(z)} \; .
\lab{phi-tilde-prime}
\ee
\item
Eq.\rf{J-nonconserv-FLRW} is equivalent to:
\be
\frac{d}{dz}\rho_0(z) - \frac{3}{1+z}\rho_0(z) 
+ \frac{d}{dz} U\bigl(\wti{\phi}(z)\bigr) = 0 \; ,
\lab{J-nonconserv-FLRW-z}
\ee
with a solution corresponding to \rf{rho-0-FLRW}:
\be
\rho_0(z) = \frac{c_0}{a_0^3} (1+z)^3 
- (1+z)^3 \int^z d\z (1+\z)^{-3} \frac{d}{d\z} U\bigl(\wti{\phi}(\z)\bigr) \; .
\lab{rho-0-FLRW-z}
\ee
\item
The Friedmann Eqs.\rf{fried-1}-\rf{fried-2} are equivalent to:
\br
6 H^2(z) = \rho_0(z) + 2M + U\bigl(\wti{\phi}(z)\bigr) \; ,
\lab{fried-1-z} \\
\frac{d}{dz} H^2(z) = \frac{2}{1+z} \rho_0(z) \; ,
\lab{fried-2-z}
\er
with $\rho_0(z)$ as in \rf{rho-0-FLRW-z}.
\end{itemize}

For the sake of confronting the observational data, Eq.\rf{fried-1-z} may be rewritten 
in terms of the various density $\Omega$-parameters:
\br
H^2(z) = H_0^2 \Bigl\lb \Omega_{dm}(z) + \Omega^{(0)}_{\L} + \Omega^{(1)}_{\L}(z)
\nonu \\
+ \Omega^{(0)}_r (1+z)^4 \Bigr\rb \; ,
\lab{fried-1-omega}
\er
where $\Omega_{dm}(z)$ stands for the ``darkon'' dark matter density parameter:
\be
\Omega_{dm}(z) \equiv \frac{\rho_0(z)}{6H_0^2} \; ,\;
\Omega_{dm}'(z) - \frac{3}{1+z}\Omega_{dm}(z) + \Omega^{(1)'}_{\L} (z)
\; ;
\lab{dm-omega}
\ee
for the dark energy density parameter:
\be
\Omega^{(0)}_{\L}\equiv \frac{2M}{6H_0^2} \quad ,\quad    
\Omega^{(1)}_{\L}(z)\equiv \frac{U\bigl(\wti{\phi}(z)\bigr)}{6H_0^2} \; ;
\lab{dlambda-omega}
\ee
and where also the contributions of radiation $\Omega^{(0)}_r$ and baryon matter 
$\Omega^{(0)}_b$ have been added.

Let us recall that the presence of the ``darkon'' potential $U$ breaks the
hidden nonlinear ``darkon'' Noether symmetry \rf{J-nonconserv} (or
\rf{J-nonconserv-FLRW} within the FLRW framework) embodying the
$\Lambda$CDM character of the original ``darkon'' model \rf{NRVF-2}, so that the 
appearance of $\Omega^{(1)}_{\L}(z)$ in \rf{fried-1-omega}-\rf{dm-omega} signifies 
deviation from $\Lambda$CDM.

%%%%%%%%%%%%%%%%%%%%%%%%%%%%%%%%%%%%%%%%%%%%%%%%%%%%%%%%%%%%%%%%%%%%%%%%%%%%%
%%%%%%%%%%%%%%%%%%%%%%%%%%%%%%%%%%%%%%%%%%%%%%%%%%%%%%%%%%%%%%%%%%%%%%%%%%%%%%%%%%%
%%%%%%%%%%%%%%%%%%%%%%%%%%%%%%%%%%%%%%%%%%%%%%%%%%%%%%%%%%%%%%%%%%%%%%%%%%%%%%%%%%%
\section{Perturbations}
\label{sec:pert}

Let us now consider scalar perturbations of the FLRW metric \rf{FLRW-metric} 
(in Newtonian gauge):
\begin{equation}
ds^2 = - (1 + 2 \Psi) dt^2 + a(t)^2 (1 - 2 \Psi) \delta_{ij} dx^i dx^j \; ,
\end{equation}
together with  perturbarions of the fields:
\be
\phi = {\wti\phi} (t) + \d\phi (t,\vec{x}) \;\; ,\;\;
 \chi = {\wti\chi} (t) + \d\chi (t,\vec{x}) \; ,
\lab{phi-chi-perturb}
\ee
where ${\wti\phi}$ and ${\wti\chi}$ are the unperturbed (``background'')
solutions for $\phi$ and $\chi$ from Eqs.\rf{phi-sol}-\rf{J-conserv-FLRW}, 
as well as perturbations of the energy density and pressure:
\be
\rho = {\wti\rho} (t) + \delta\rho (t,\vec{x}) \;\; ,\;\;
p = {\wti p} (t) + \delta p (t,\vec{x}) \; ,
\lab{rho-p-perturb}
\ee
where ${\wti\rho}$ and ${\wti p}$ are the unperturbed background values 
of $\rho$ and $p$ in \rf{fried-1} and \rf{fried-2}.
% In addition, we will introduce a ``darkon'' Noether symmetry breaking
% perturbation $U(\phi) \equiv \d U(\phi)$ -- cf. Eqs.\rf{J-nonconserv}
% and \rf{fluid-rho-p}-\rf{rho-0}, so that:
Explicitly:
\br
\d\rho = \d\rho_0 + U^\pr\,\d\phi \quad ,\quad \d p = - U^\pr\,\d\phi \; .
\lab{delta-rho-p}\\
\d\rho_0 = \rho_0 \frac{\d\chi}{1+{\wti\chi}} 
+ 2 (1+{\wti\chi})V^\pr\,\d\phi \; .
\lab{delta-rho-0}
\er
% cf. Eqs.\rf{fluid-rho-p}-\rf{rho-0}.

% Perturbation of \rf{fluid-velocity}:
% \begin{equation}
% u_{\mu} = - \frac{\partial_\mu \phi}{\sqrt{2X}} = 
% (- 1,\vec{0}) + (- \Psi, \frac{\vec{\nabla} \delta\phi}{\phi})
% \end{equation}
The perturbation of fluid velocity unit vector \rf{fluid-velocity} reads:
\be
\d u_\m = (-\Psi, \d u_i) \;\; ,\;\; \d u_i = - \frac{\pa_i \d\phi}{\phidot}
\lab{u-perturb}
\ee

The perturbation of the dynamical constraint Eq.\rf{L-const} around the FLRW
background::
\br
% \frac{d}{dt}\Bigl(\dot{\phi} \delta\dot\phi - \dot{\phi}^2 \Psi 
% - V'(\phi) \d\phi\Bigr) = 0 \quad {\rm i.e.},
%\lab{L-const-perturb-0} \\
\dot{\phi} \delta\dot\phi - \dot{\phi}^2 \Psi 
- V'(\phi) \d\phi = 0   %\d M \; ,
\lab{L-const-perturb}
\er
%  where $\d M$ is a small constant, 
or, equivalently using \rf{phiddot}:
\be
\d\phidot = \phidot \Psi + \frac{\phiddot}{\phidot} \d\phi
\lab{delta-phi-eq}
\ee
yields solution for $\d\phi (t,\vec{x})$:
% \be
% \d\phi (t,\vec{x}) = % \sqrt{2(V(\phi(t))-2M}\Bigl\lb
% \phidot \Bigl\lb \int dt^\pr \Bigl(\Psi (t^\pr,\vec{x}) 
% + \frac{\d M}{\phidot^2}\Bigr) + C_0 (\vec{x})\Bigr\rb \; 
% \lab{dphi-sol-0}
% \ee
% with $C_0 (\vec{x})$ some infinitesimal function of the spacelike coordinates.
% We now observe that, at least for the choice \rf{V-0} of the scalar
% potential $V$, the presence of a non-zero $\d M$ will create a perturbative
% instability -- a linearly rising with $t$ term
% $\frac{\d M}{\phidot}\,t$ in $\d\phi$ \rf{dphi-sol-0}.
% % Note that the expresion \rf{dphi-sol} for $\d\phi$ does not depend on the
% %form of the potential $V(\phi)$.
% Thus, $\d M$ in \rf{L-const-perturb} must be zero, and Eq.\rf{dphi-sol-0} 
% acquires the form:
\be
\d\phi (t,\vec{x}) = % \sqrt{2(V(\phi(t))-2M}\Bigl\lb
\phidot \Bigl\lb \int dt^\pr \Psi (t^\pr,\vec{x}) + C_0 (\vec{x})\Bigr\rb \; 
\lab{dphi-sol}
\ee
with $C_0 (\vec{x})$ some infinitesimal function of the spacelike coordinates.

The perturbations of the stress-energy tensor components \rf{EM-tensor} read:
\br
% \d T_{00}=\bigl(\rho_0 + 2M)2\Psi + \d\rho \quad,\quad
\d T^0_0 = - \d\rho = - \d\rho_0 - U^\pr\,\d\phi \; ,
\lab{delta-T00} \\
% \d T_{0i}= -\rho_0 \d u_i = \rho_0 \frac{\pa_i \d\phi}{\phidot} \quad ,\quad 
\d T^i_0 = - \frac{1}{a^2}\rho_0 \d u_i 
= \frac{1}{a^2}\rho_0 \frac{\pa_i \d\phi}{\phidot} \; ,
\lab{delta-Ti0} \\
% \d T_{ij} = a^2 \d_{ij} \bigl\lb 4M \Psi + \d p \quad ,\quad
\d T^i_j = - \d^i_j \d p = \d^i_j U^\pr\,\d\phi \; .
\lab{delta-Tij}
\er
Let us now consider the zeroth component of the perturbed energy-momentum
conservation equation (cf. \textsl{e.g.} \cite{weinberg-2008}):
\br
\pa_0 \d T^0_0 +\pa_i \d T^i_0 + 3H \d T^0_0 - H \d T^i_i
\nonu \\
- \frac{\rho_0}{2a^2} \Bigl(\frac{d}{dt}\d g_{ii} - 2H \d g_{ii}\Bigr) = 0 \; ,
\lab{EM-perturb-conserv-0}\\
\d g_{ij} \equiv - 2 \Psi \,a^2 \d_{ij} \; ,
\nonu
\er
which upon inserting \rf{delta-T00}-\rf{delta-Tij} becomes:
\be
\Bigl(\frac{d}{dt}+3H\Bigr)\d\rho + 3H \d p  
% + \frac{\rho_0}{a^2} \nabla_i \d u_i -3 \rho_0 \Psidot = 0
- \frac{\rho_0}{a^2\,\phidot} \nabla^2 \d\phi - 3\rho_0 \Psidot = 0
\; .
\lab{EM-perturb-conserv}
\ee
% or, equivalently:
% \br
% \Bigl(\frac{d}{dt}+3H\Bigr)\d\chi - 3(1+\chi)\Psidot 
% \nonu \\
% - \frac{(1+\chi)}{a^2\,\phidot} \nabla^2 \d\phi + \frac{U^\pr}{\phidot} \Psi
% + \frac{U^{\pr\pr}}{\phidot} \d\phi = 0 \; .
% \lab{EM-perturb-conserv-1}
% \er

% On the other hand, the space components of the perturbed energy-momentum
% conservation equation yield:
% \be
% \bigl(\frac{d}{dt}+3H\Bigr) \bigl(\rho_0 \d u_i\bigr) + \pa_i \d p  
% + \rho_0 \pa_i\Psi = 0 
% \lab{EM-perturb-conserv-i}
% \ee
% or, more explicitly: % using \rf{u-perturb}, \rf{p-perturb} and \rf{rho-0-FLRW}:
% \be
% \bigl(\frac{d}{dt}+3H\Bigr) \bigl(\rho_0 \frac{\d\phi}{\phidot}\bigr)
% + U^\pr \d\phi - (1+\chi)\phidot^2 \Psi = 0 \; .
% \lab{EM-perturb-conserv-i-1}
% \ee
% The latter is automatically satified upon using \rf{rho-0-eq}. 
%%%%%%%%%%%%%%%%%%%%%%%%%%%%%%%%%%%%%%%%

Introducing the dark matter energy density contrast:
\be
\d_{DM} \equiv \frac{\d\rho_0}{\rho_0}
\lab{DM-contrast}
\ee
and using Eq.\rf{EM-perturb-conserv} by taking into account
\rf{J-nonconserv-FLRW} and last Eq.\rf{delta-rho-p} we obtain:
\be
\frac{d}{dt} \d_{DM} - \frac{\nabla^2 \d\phi}{a^2\,\phidot} - 3\Psidot
- \frac{\Udot}{\rho_0}\,\d_{DM}
+ \frac{1}{\rho_0}\frac{d}{dt}\bigl(U^\pr\,\d\phi\bigr) = 0 \; .
\lab{DM-contrast-eq-1}
\ee
Applying time-derivative $\frac{d}{dt}$ on Eq.\rf{DM-contrast-eq-1} and
using Eq.\rf{delta-phi-eq} -- specific perturbation equation for the present 
``darkon'' model of dynamical dark matter, 
as well as using one of the perturbed Einstein
equations for the metric perturbation component $\Psi$ 
(see \textsl{e.g.} \cite{baumann}):
\be
\frac{1}{a^2}\nabla^2 \Psi = 
\frac{1}{4}\Bigl(\d\rho_0 + U^\pr\,\d\phi - 3aH\rho_0\d\phi\Bigr) \; ,
\lab{einstein-perturb}
\ee
we obtain the second-order differential equation for the dark matter contrast:
\br
\frac{d^2}{dt^2} \d_{DM} + 2 H \frac{d}{dt} \d_{DM} + \frac{1}{4} \rho_0 \d_{DM}
\nonu \\
- 3\Bigl\lb\ddot{\Psi} + 2 H \dot{\Psi} - \frac{1}{4} aH \rho_0 \d\phi\Bigr\rb
\nonu \\
=  \frac{1}{4} U^\pr\,\d\phi - \Bigl(\frac{d}{dt} + 2 H \Bigr)
\Bigl\lb \frac{1}{\rho_0}\frac{d}{dt}\bigl(U^\pr\,\d\phi\bigr) 
- \frac{\Udot}{\rho_0}\d_{DM}\Bigr\rb \; .
\lab{DM-contrast-eq-2}
\er
Recall that $\rho_0$ and $\d\phi$ are explicitly given by \rf{J-nonconserv-FLRW} and 
\rf{dphi-sol}, respectively. 

In the case $U(\phi)=0$ (or $U(\phi)={\rm const}$) when the ``darkon'' 
nonlinear Noether symmetry is intact \rf{J-conserv}, the r.h.s. of
Eq.\rf{DM-contrast-eq-2} vanishes and it reduces to:
\br
\begin{split}
\frac{d^2}{dt^2} \d_{DM} + 2 H \frac{d}{dt} \d_{DM} + \frac{1}{4} \rho_0 \d_{DM} \\- 3\Bigl\lb\ddot{\Psi} + 2 H \dot{\Psi} 
- \frac{1}{4} Ha \rho_0\d\phi\Bigr\rb = 0 \; ,
\lab{DM-contrast-relativistic}    
\end{split}
\er
where $\rho_0$ is now given by \rf{J-conserv-FLRW} and $\d\phi$ is expressed
through the metric perturbation $\Psi$ according to \rf{dphi-sol}. 
Eq.\rf{DM-contrast-relativistic} is the general relativistic form of the
equation for the dark matter density contrast over $\Lambda$CDM FLRW background.
In the subhorizon limit where the metric perturbation $\Psi$ is small
\cite{baumann} the terms in the square brackets on the l.h.s. of 
\rf{DM-contrast-relativistic} can be ignored, so that the latter 
% dark energy density contrast Eq.\rf{DM-contrast-eq-2}
simplifies to the familiar form of the equation for the energy 
density contrast of generic dark matter perturbations on $\Lambda$CDM
background in the Newtonian limit \cite{baumann} (recall, we are using units with
$16\pi G_{\rm Newton} = 1$):
\be
\frac{d^2}{dt^2} \d_{DM} + 2 H \frac{d}{dt} \d_{DM} 
+ \frac{1}{4} \rho_0 \d_{DM}=0 \; .
\lab{DM-contrast-nonrel}
\ee

%%%%%%%%%%%
In terms of redshift $z$ Eq.\rf{DM-contrast-eq-2} takes the form:
\br
\d^{\pr\pr}_{DM} + \d^{\pr}_{DM}\Bigl(\frac{H^\pr(z)}{H(z)} - \frac{1}{1+z}\Bigr)
+ \frac{\rho_0(z) \d_{DM}}{4(1+z)^2 H^2(z)} 
\nonu \\
= \frac{d}{dz}\Bigl(\frac{U^\pr(z)}{\rho_0(z)}\d_{DM}\Bigr) 
+ \frac{U^\pr(z)}{\rho_0(z)}\d_{DM} \Bigl(\frac{H^\pr(z)}{H(z)} - \frac{1}{1+z}\Bigr)
\; ,
\lab{DM-contrast-z}
\er
with primes indicating $\frac{d}{dz}$ and where $\rho_0(z), H^2 (z), H^\pr (z)$ 
are to be replaced by the expressions \rf{rho-0-FLRW-z}, \rf{fried-1-z} and
\rf{fried-2-z}, respectively.
Here again, as in \rf{DM-contrast-nonrel} above, the subhorizon approximation (Newtonian
limit) \cite{baumann} was used (\textsl{i.e.}, the terms involving the metric
perturbation $\Psi$ are ignored). 

%%%%%%%%%%
Let us recall that the growth rate function is definded as: 
\begin{equation}
f\equiv \frac{d\ln\delta}{d\ln a}\quad \text{or} \quad f\equiv\frac{\delta'}{\delta},
\end{equation}
with $\d = \d\rho/\rho$ denoting the pertinent matter density contrast, which depicts 
how quickly the perturbations evolve. 
Typically, observational data on the growth of structure are presented as constraints 
on the parameter
\begin{equation}
f\sigma_8 (z) = - (z+1) \sigma_8(0) \frac{\delta'(z)}{\delta(0)},
\end{equation}
which can directly be extracted from redshift space distortion data. 
The $\sigma_8(0)$ is the present amplitude of the matter power spectrum at the 
scale of $8h^{-1}$Mpc \cite{Raccanelli:2015qqa,Macaulay:2013swa}.
%%%%%%%%%%%%%%%%%%%%%%%%%%%%%%%%%%%%%%%%%%%%%%%%%%%%%%%%%%%%%%%%%%%%%%%%%%%%%

%%%%%%%%%%%%%%%%%%%%%%%%%%%%%%%%%%%%%%%%%%%%%%%%%%%%%%%%%%%%%%%%%%%%%%%%%%%%%

%%%%%%%%%%%%%%%%%%%%%%%%%%%%%
\begin{figure*}[t!]
\centering
\includegraphics[width=0.8\textwidth]{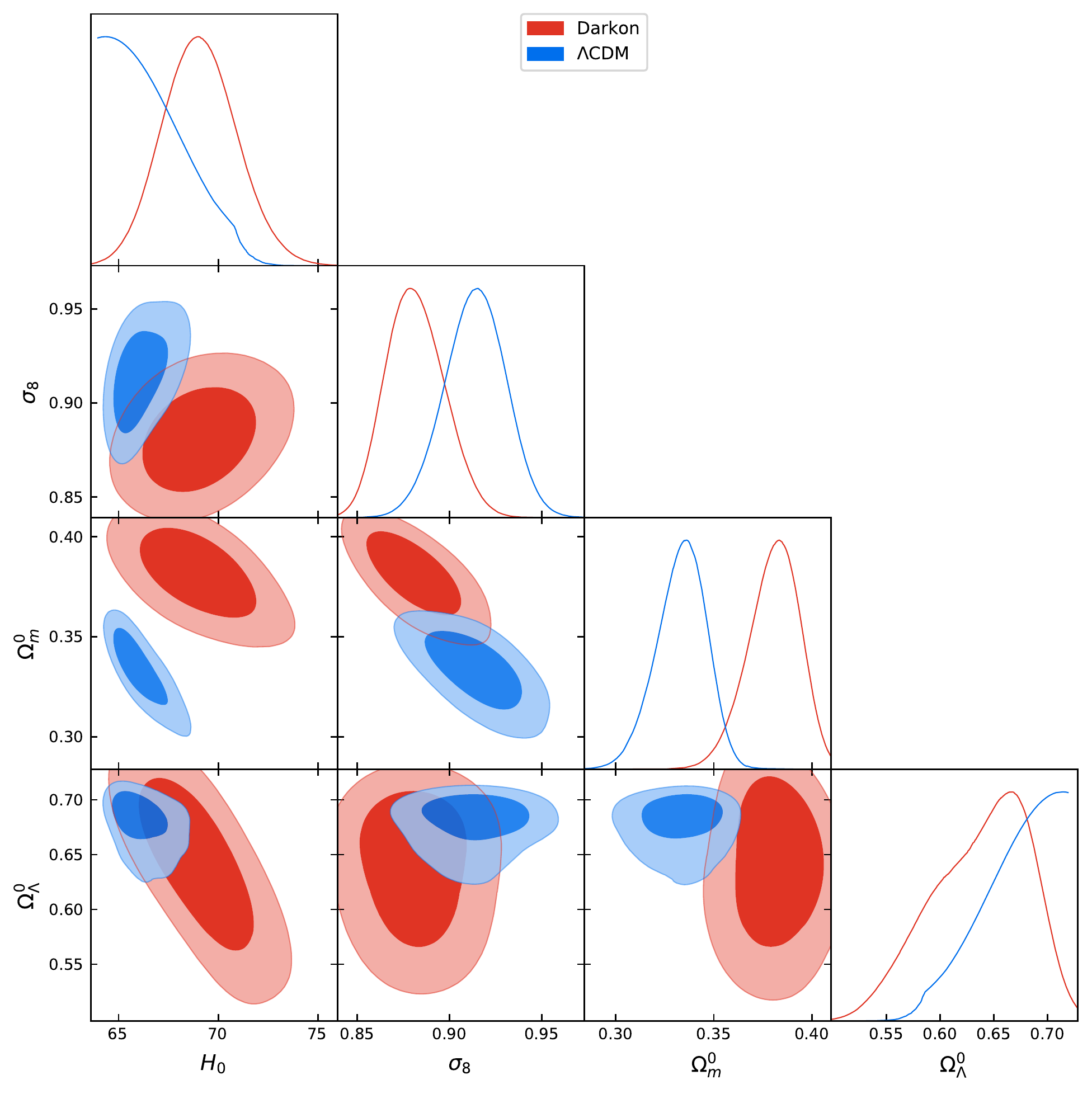}
\caption{{\it{The corner plot of the potential (\ref{eq:potential}) 
with the direct measurements of the Hubble expansion and the growth matter data.}} }
\label{fig:cpl}
\end{figure*}
%%%%%%%%%%%%%%%%%%%%%%%%%%%%%
\begin{table*}
\tabcolsep 5.5pt
\vspace{1mm}
\begin{tabular}{cccccccc} 
\hline \hline
Potential & $\Omega_{m}^{0}$ & $H_0 \frac{km/s}{Mpc})$ & $\Omega_{\Lambda}^{0}$  & $ \beta$ & $\sigma_8$ &
$\chi^{2}$  \vspace{0.05cm}
\\ \hline
%----------------------------------------
\hline
  (\ref{eq:potential}) & $0.38\pm 0.013$ & $69.04_{\pm 1.75}$ & $0.638_{\pm 0.04}$ & $0.46_{\pm 0.03}$ & $0.882_{\pm 0.178}$ & 84.56  
\\
Flat ($\Lambda$CDM) &  $0.334_{\pm 0.012}$ & $66.07_{\pm0.877}$ & $0.6796_{\pm 0.0194}$ & $0$ & $0.913\pm_{0.015}$ & 96.12
\\

\hline\hline
\end{tabular}
\caption[]{ \it{Observational constraints and the
corresponding $\chi^{2}$ for the considered cosmological models.}}
\label{tab:Results}
\end{table*}
%%%%%%%%%%%%%%%%%%%%%%%%%%%%%%%%%%%%%%%%%

\section{Statistical Analysis}
\label{sec:Data}
In order to assess the viability of the model, we confront it with the observational 
data the solutions for 
Eq.\rf{fried-1-omega} (the homogeneous one within the FLRW framework) 
and Eq.\rf{DM-contrast-eq-2} (for the perturbations above the FLRW background).  

We examine the following ``darkon'' Noether
symmetry-breaking potential (with $\tilde{\phi}$ -- the redefined ``darkon''
field \rf{phi-tilde}):
\begin{equation}
U(\tilde{\phi}) = 2M \beta^2  \tilde{\phi} \; .
\label{eq:potential}
\end{equation}
For the limit $\beta \rightarrow 0$ the potential goes to zero, and we recover the 
$\Lambda$CDM model both in the homogeneous solution as well as on the linear 
perturbation level. 

We test the solutions that are provided by the present ``darkon'' model with two data 
sets: the direct measurements of the Hubble expansion\cite{Yu_2018,Moresco_2018} and the growth rate data set 
\cite{Sagredo:2018ahx,Anagnostopoulos:2019miu,Basilakos:2019hlb,Kazantzidis:2018rnb,Gannouji:2018ncm,Kazantzidis:2018jtb,Benisty:2020kdt}.

The direct measurements of the Hubble expansion set contains $N=36$ measurements of the Hubble expansion 
in the redshift range $0.07\leq z\leq 2.33$. 5 measurements are based on 
Baryonic Acoustic Oscillations (BAOs), and the other estimated via the 
differential age of passive evolving galaxies. 
Here, the corresponding $\chi^2_{H}$ function reads:
\begin{equation}
\chi_{\mathcal{H}}^{2}={\bf \cal H}\,
{\bf C}_{H,\text{cov}}^{-1}\,{\bf \cal H}^{T}\,,
\end{equation}
where  ${\bf \cal H}=\{H_{1}-H_{0}E(z_{1},\phi^{\nu})\,,\,...\,,\, H_{N}-H_{0}E(z_{N},\phi^{\nu})\}$ 
and $H_{i}$ are the observed Hubble rates at redshift $z_{i}$ ($i=1,...,N$). 
The matrix ${\bf C}$ denotes the covariance matrix, 
and $\phi^{\nu}$ denotes the other parameters on which the Hubble rate depends.

A model-independent cosmological probe, the $f \sigma_{8}$ product, is estimated from the analysis of redshift-space distortions \cite{Song:2008qt,Benisty:2020kdt}. There is a big number of data points. We choose to use a compilation of $f\sigma_8$ data that checked in terms of its robustness using information theoretical methods. The relevant chi-square function reads
\begin{equation}
    \chi^2_{f\sigma 8} ={f\sigma 8}\,
{\bf C}_{f\sigma 8,\text{cov}}^{-1}\,{f\sigma 8}^{T}\,,
\end{equation}
where $f\sigma_{8}(a_{i},\phi^{\nu+1})_{theor}= 
\sigma_{8}\delta'(a_{i},\phi^{\nu})/\delta(1,\phi^{\nu})a_{i}$ and a prime 
denotes derivative of the scale factor $a$ with the corresponding correlation matrix. The quantity $\sigma_{8}$ is a free parameter. The statistical vector $\phi^{\nu}$  
contains the other free parameters of the statistical model. The values $\delta'(a_{i})$, $\delta(1)$ are calculated by the numerical solution of Eq. 
Eq.\rf{DM-contrast-eq-2} for a given set of cosmological parameters. 

To obtain the joint constraints on the cosmological parameters from 2 cosmological 
probes, we define the total $\chi^2_{\text{tot}}$ expression:
        \begin{equation}
        \chi_{\text{tot}}^2 = \chi_{\mathcal{H}}^2 + \chi^2_{f\sigma 8}\,.
        \end{equation}
Regarding the problem of data fit, we
use a nested sampler as it is implemented within the open-source $Polychord$
\cite{Handley:2015fda} with the $GetDist$ packaged \cite{Lewis:2019xzd} to present the results. The prior we choose is with a uniform distribution, where $\Omega_{m} \in [0.2;0.4]$, $\Omega_{\Lambda}\in[0.5;0.8]$, $\sigma_8\in [0.5;1.2]$, $H_0\in [65;75]$ and $\beta \in [0;0.5]$ 
for the ``darkon'' model.

Fig. \ref{fig:cpl} presents the corner plot of the joint statistical
analyses. Table \ref{tab:Results} summarizes the joint statistics. One can
see that the $\sigma_8$ that the potential \rf{potential} predicts 
is closer to the value predicted by 
PLANCK collaboration $\sigma_8 = 0.811 \pm 0.006$ and $H_0 = 67.4 \pm 0.5  km/s/Mpc$. 
%Therefore the present model is a good candidate for description of the 
%dynamical dark energy and dark matter and the related effect of the tension 
%between $\sigma_8$ values. %However, the $\chi^2$ is larger then the $\Lambda$CDM $\chi^2$ values. Therefore additional test should be done in the future research to test the viability of the model.
The $\chi^2$ per degrees of freedom for the ``darkon'' model yields $1.023$, while for $\Lambda$CDM the $\chi^2$ per degrees of freedom gives $0.899$. The fit is better when the Noether Symmetry is preserve.  

%%%%%%%%%%%%%%%%%%%%%%%%%%%%%%%%%%%%%%%%%%%%%%%%%%%%%%%%%%%%%%%%%%%%%%%%%%%
%%%%%%%%%%%%%%%%%%%%%%%%%%%%%%%%%%%%%%%%%%%%%%%%%%%%%%%%%%%%%%%%%%%%%%%%%%%%%
\section{Conclusions}
\label{sec:conclusion}
This paper connects the standard $\Lambda$CDM  model of cosmology to 
the hidden nonlinear Noether symmetry of a simple modified 
gravity-matter model with a single scalar field based on the formalism of 
non-Riemannian spacetime volume-elements. Via the Noether symmetry of its 
modified action the scalar field, called ``darkon'', dynamically generates
both cosmological constant (not present in the original action), as well as
dust-like dark matter component of the pertinent stress-energy tensor -- a
simplest explicit realization of the $\Lambda$CDM framework.
Adding Noether symmetry-breaking ``darkon'' potential introduces interaction
(energy transfer) between dark energy and dark matter.

We calculate up to linear order of perturbations the solution for the above theory 
confirming that in the absence of ``darkon'' Noether symmetry breaking the
known equation for the dark matter density contrast for the $\Lambda$CDM
scenario is recovered.

We also studied the homogeneous background and linearly perturbed solutions 
with a specific plausible choice of ``darkon'' Noether symmetry-breaking potential.
Using the direct measurements of the Hubble expansion and the growth matter perturbations data we find that our fit is closer to Planck data, and the $\chi^2/Dof$ is not significantly higher. However, to alleviate the cosmic tensions completely we should test more data sets as pantheon Type Ia supernova and measurements from the early universe as the CMB data. 

%%%%%%%%%%%%%%%%%%%%%%%%%%%%%%%%%%%%%%%%%%%%%%%%%%%%%%%%%%%%%%%%%%%%%%%%%%%%%%
%%%%%%%%%%%%%%%%%%%%%%%%%%%%%%%%%%%%%%%%%%%%%%%%%%%%%%%%%%%%%%%%%%%%%%%%%%%%%%
\acknowledgments
We all are grateful for support by COST Action CA-15117 (CANTATA), COST Action 
CA-16104 and COST Action CA-18108. D.B. thanks Ben-Gurion University of the Negev and 
Frankfurt Institute for Advanced Studies for generous support. E.N. and S.P. are partially supported by Bulgarian National Science Fund Grant DN 18/1.
We also thank the referee for useful remarks.
%%%%%%%%%%%%%%%%%%%%%%%%%%%%%%%%%%%%%%%%%%%%%%%%%%%%%%%%%%%%%%%%%%%%%%%%%%%%%
\bibliographystyle{apsrev4-1}
\bibliography{ref}

\end{document}